\documentclass[prb,aps, twocolumn, amsmath,amssymb,superscriptaddress]{revtex4}
\usepackage{graphicx} 
\usepackage{float}
\usepackage{amsmath}
\usepackage{amssymb}
\usepackage{amsfonts}
\usepackage{bm}
\usepackage{euscript}
\usepackage{enumerate}
\usepackage{hhline}
\usepackage{pslatex}
\usepackage{tabularx}
\usepackage[usenames,dvipsnames]{xcolor}

\usepackage[colorlinks,linkcolor=red,anchorcolor=blue,citecolor=blue]{hyperref}

\makeatletter
\renewcommand{\@biblabel}[1]{#1. }
\renewcommand{\@dotsep}{500}
\renewcommand{\@pnumwidth}{0em}
\renewcommand{\l@figure}[2]{
	\@dottedtocline{1}{1.5em}{2em}{Figure #1}{}\vspace{15pt}}

\begin{document}
	
	\title{High-performance Sources of Multidimensionally Engineered Quantum Light Based on Monolithic Microcavity-metalens Interfaces}
	
	\author{Jiantao Ma}
	\thanks{These authors contributed equally to this work.}
	\affiliation{State Key Laboratory of Optoelectronic Materials and Technologies, School of Physics, School of Electronics and Information Technology, Sun Yat-sen University, Guangzhou, 510275, China.}
	
	\author{Dong Liu}
	\thanks{These authors contributed equally to this work.}
	\affiliation{State Key Laboratory of Optoelectronic Materials and Technologies, School of Physics, School of Electronics and Information Technology, Sun Yat-sen University, Guangzhou, 510275, China.}
	
	\author{Shunfa Liu}
	\thanks{These authors contributed equally to this work.}
	\affiliation{State Key Laboratory of Optoelectronic Materials and Technologies, School of Physics, School of Electronics and Information Technology, Sun Yat-sen University, Guangzhou, 510275, China.}
	
	\author{Jiawei Yang}
	\thanks{These authors contributed equally to this work.}
	\affiliation{State Key Laboratory of Optoelectronic Materials and Technologies, School of Physics, School of Electronics and Information Technology, Sun Yat-sen University, Guangzhou, 510275, China.}
	
	\author{Nilo Mata-Cervera}
	\affiliation{Centre for Disruptive Photonic Technologies, School of Physical and Mathematical Sciences, Nanyang Technological University, Singapore 637371, Singapore.}
	
	\author{Bo Chen}
	\affiliation{State Key Laboratory of Optoelectronic Materials and Technologies, School of Physics, School of Electronics and Information Technology, Sun Yat-sen University, Guangzhou, 510275, China.}
	
	\author{Xueshi Li}
	\affiliation{State Key Laboratory of Optoelectronic Materials and Technologies, School of Physics, School of Electronics and Information Technology, Sun Yat-sen University, Guangzhou, 510275, China.}
	
	\author{Guixin Qiu}
	\affiliation{State Key Laboratory of Optoelectronic Materials and Technologies, School of Physics, School of Electronics and Information Technology, Sun Yat-sen University, Guangzhou, 510275, China.}
	
	\author{Kaixuan Chen}
	\affiliation{Guangdong Provincial Key Laboratory of Optical Information Materials and Technology, South China Academy of Advanced Optoelectronics, South China Normal University, Guangzhou, 510006, China.}
	
	\author{Hanqing Liu}
	\affiliation{State Key Laboratory for Superlattices and Microstructures, Institute of Semiconductors, Chinese Academy of Sciences, Center of Materials Science and Optoelectronics Engineering, University of Chinese Academy of Sciences, Beijing, 100083, China.}
	
	\author{Haiqiao Ni}
	\affiliation{State Key Laboratory for Superlattices and Microstructures, Institute of Semiconductors, Chinese Academy of Sciences, Center of Materials Science and Optoelectronics Engineering, University of Chinese Academy of Sciences, Beijing, 100083, China.}
	
	\author{Dunzhao Wei}
	\affiliation{State Key Laboratory of Optoelectronic Materials and Technologies, School of Physics, School of Electronics and Information Technology, Sun Yat-sen University, Guangzhou, 510275, China.}

	\author{Zhichuan Niu}
	\affiliation{State Key Laboratory for Superlattices and Microstructures, Institute of Semiconductors, Chinese Academy of Sciences, Center of Materials Science and Optoelectronics Engineering, University of Chinese Academy of Sciences, Beijing, 100083, China.}
	
	\author{Ying Yu}
	\affiliation{State Key Laboratory of Optoelectronic Materials and Technologies, School of Physics, School of Electronics and Information Technology, Sun Yat-sen University, Guangzhou, 510275, China.}
	
	\author{Yijie Shen}
	\thanks{yijie.shen@ntu.edu.sg}
	\affiliation{Centre for Disruptive Photonic Technologies, School of Physical and Mathematical Sciences, Nanyang Technological University, Singapore 637371, Singapore.}%
	\affiliation{School of Electrical and Electronic Engineering, Nanyang Technological University, Singapore 639798, Singapore}
	
	\author{Liu Liu}
	\thanks{liuliuopt@zju.edu.cn}
	\affiliation{State Key Laboratory for Modern Optical Instrumentation, College of Optical Science and Engineering, International Research Center for Advanced Photonics, Zhejiang University, Hangzhou, 310058, China.}
	
	\author{{Xuehua Wang}}
	\thanks{ wangxueh@mail.sysu.edu.cn}
	\affiliation{State Key Laboratory of Optoelectronic Materials and Technologies, School of Physics, School of Electronics and Information Technology, Sun Yat-sen University, Guangzhou, 510275, China.}
	\affiliation{Quantum Science Center of Guangdong–Hong Kong–Macao Greater Bay Area, Shenzhen, 518105, China.}
	
	\author{{Jin Liu}}
	\thanks{liujin23@mail.sysu.edu.cn}
	\affiliation{State Key Laboratory of Optoelectronic Materials and Technologies, School of Physics, School of Electronics and Information Technology, Sun Yat-sen University, Guangzhou, 510275, China.}
	\affiliation{Quantum Science Center of Guangdong–Hong Kong–Macao Greater Bay Area, Shenzhen, 518105, China.}
	
	\begin{abstract}
		\noindent \textbf{The ultimate non-classic light sources for modern photonic quantum technology require on-demand generation of indistinguishable quantum light with high brightness and flexible engineering of quantum emission in multiple degrees of freedom. In this work, we present monolithic microcavity-metalens interfaces consisting of quantum-dot-micropillar single-photon sources and ultra-thin metalenses accurately aligned on opposite sides of an III-V compound semiconductor chip. The pronounced cavity quantum electrodynamics effect enabled by the micropillar cavity facilitates single-photon emission from quantum dots with simultaneous high degrees of single-photon purity, source brightness and photon indistinguishability while the multi-functional metalenses concurrently tailor quantum emission in multiple physical degrees of freedom including radiation divergence, emission directionality, polarization state and orbital angular momentum (OAM). Furthermore, high-fidelity polarization-OAM entanglement and single photons with local spin topologies are successfully generated in our integrated device. In particular, we demonstrate stable propagations of single-photon skrymions in atmospheric
			turbulence and reveal their topological advantages over the conventional structured quantum light. Our work advances the research fields of integrated quantum photonics and meta-optics, providing integrated high-dimensional quantum light sources for advanced photonic quantum science and technology.}
	\end{abstract}
	
	\maketitle   
		
		Quantum radiation of light serves as a cornerstone for modern photonic quantum technology\cite{o2009photonic} with applications across quantum communication\cite{gisin2007quantum}, quantum computing\cite{o2007optical,zhong2020quantum,madsen2022quantum} and quantum metrology\cite{giovannetti2011advances}. As the most prominent quantum state of light, single photons can be deterministically generated in two-level systems such as natural atoms\cite{mckeever2004deterministic}, trapped ions\cite{keller2004continuous} and solid-state quantum emitters\cite{aharonovich2016solid}. Among them, solid-state quantum emitters, e.g., excitons in quantum dots (QDs)\cite{michler2017quantum,senellart2017high,zhou2023epitaxial}, defects in wide bandgap semiconductors\cite{castelletto2014silicon,zhou2018room} and layered materials\cite{montblanch2023layered}, are particularly appealing for integrated quantum photonic technology\cite{,wang2020integrated,elshaari2020hybrid} due to their compatibility with the modern nanofabrication process. To date, epitaxial QDs are leading the device performance of quantum light sources in terms of low multi-photon probability, record source brightness, high entanglement fidelity and near-unity photon indistinguishability\cite{somaschi2016near,he2017deterministic,wang2019towards,liu2019solid,wang2019demand,uppu2020scalable,tomm2021bright,wei2022tailoring} by harnessing the accelerated spontaneous emission rate and improved emission directionality empowered by the cavity quantum electrodynamics (cQED) effect. However, cavity-enhanced solid-state quantum light sources predominantly operate in the simplest Gaussian-like modes in linear polarization states\cite{gazzano2013bright,liu2019solid,tomm2021bright,yang2024tunable}, which limits their applications in high-dimensional quantum information processing involving manipulation of quantum emission in multiple degrees of freedom. Alternatively, broadband nanostructures have been widely employed for molding the flow of light\cite{kan2023advances, ma2024engineering}. In particular,  metasurfaces provide an ultra-thin solution of flexibly tailoring photonic states in multiple degrees of freedom\cite{chen2016review,chen2020flat,ma2024engineering}, which have been successfully implemented in the manipulation of quantum states of light by either having the quantum sources integrated\cite{li2020metalens,huang2019monolithic,wu2022room,bao2020demand,komisar2023multiple,jia2023multichannel} or separated\cite{stav2018quantum,wang2018quantum,li2023arbitrarily} from the metasurfaces. Nevertheless, indistinguishable and bright quantum light sources have not been achieved in the metalens-quantum emitter platform due to the relatively weak light-matter interaction strength, i.e., limited Purcell effect, provided by metasurfaces. The main challenge in pursuing integrated devices capable of simultaneously generating indistinguishable single photons and multi-dimensionally engineering their optical states is the intrinsic incompatibility between the relatively sharp cavity resonances for enhancing light-matter interactions and the broadband metasurfaces for manipulating optical states. In addition, flying qubits featuring topological textures could potentially provide advantages such as robust light propagation and high-density storage. Achieving all the crucial ingredients, including source performance (brightness, single-photon purity and photon indistinguishability), mode characteristics (directionality, polarization, OAM) and topological spin textures within a universal platform is highly desirable yet technologically formidable.
		
		\begin{center}
			\begin{figure*}
				\begin{center}
					\includegraphics[width=\linewidth]{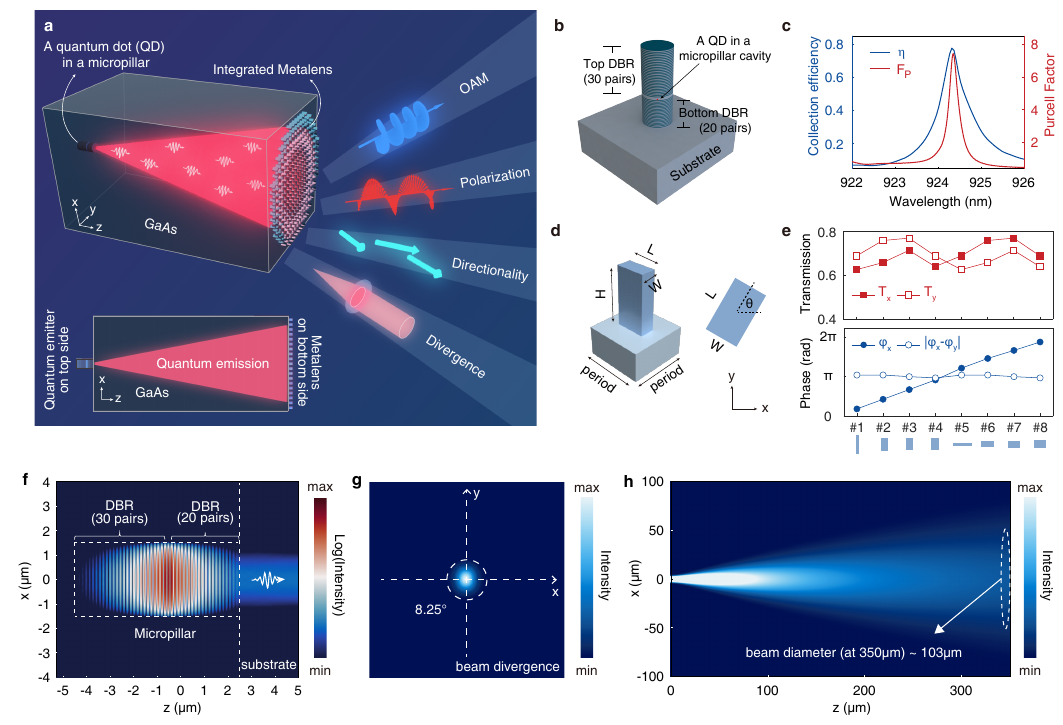}
					\caption{\textbf{Operation principle of the monolithic microcavity-metalens interfaces.}  (a) Schematics of the monolithic microcavity-metalens interface. A QD-micropillar single-photon device emits bright and indistinguishable single photons toward its substrate. A metalens on the backside of the chip arbitrarily engineers the photon states of the quantum emission in multiple physical degrees of freedom including radiation divergence, emission directionality, polarization state and OAM. Inset: the XZ plane cross-section of the device. (b) Schematics of the single-photon source on the front side of the chip, consisting of a single QD embedded in the center of a micropillar cavity. The top mirror has 10 more pairs of the DBR than the bottom mirror, which gives rise to the downward photon emission. (c) Numerical simulations of the Purcell factor and the extraction efficiency of the ideal single-photon source. (d) Schematics of the meta-atom acting as a truncated rectangular waveguide to engineer the amplitude and the phase front of the single photons. (e) The calculated transmission and phase shift for meta-atoms with different shapes. (f) Numerical simulation of the single-photon emission to the substrate. (g) Calculated beam divergence of single-photon emission at the output of the micropillar. The DBRs are fully etched in our process. (h) Numerical simulation of the beam propagation of single photons in the substrate. The beam diameter is expanded from $\sim$2~$\mu$m to $\sim$103~$\mu$m by propagating through a 350~$\mu$m-thickness substrate.}
					\label{fig:Fig1}
				\end{center}
			\end{figure*}
		\end{center}

		In this work, we experimentally demonstrate monolithic microcavity-metalens interfaces uniquely combining the advantages of large Purcell enhancements enabled by the cavity QED effect and flexible optical state manipulation empowered by metasurfaces. Due to the unique configuration of the microcavity-metalens interface, we have successfully matched the mode profiles between the 2-$\mu$m-diameter output of the micropillar and the 200-$\mu$m-diameter metalens, without compromising the cavity performance. Indistinguishable single-photon generation and on-demand photonic state engineering are achieved in an integrated device, exhibiting a multi-photon suppression with a $g^{(2)}(0)$ = 0.070(1), a photon extraction efficiency (at the first lens) up to 35.7(5)\%, a photon indistinguishability (with a delay of 13 ns) of 0.736(2) and arbitrarily tailored optical states in radiation divergence, emission directionality, polarization state and orbital angular momentum (OAM). We further demonstrate the on-demand generation of polarization-OAM entanglement with high fidelity. Finally, single photons whose spin states feature various skyrmonic textures are achieved. Such a unique combination of quantum light sources and metalenses fuses the very active research fields of integrated quantum photonics and meta-optics\cite{solntsev2021metasurfaces,ma2024engineering,ding2023advances}. The state-of-the-art solid-state quantum light sources equipped with the ability of on-demand photonic state manipulation bring unprecedented opportunities to explore structured light-matter interactions at the nanoscale and high-dimensional integrated photonic quantum technology\cite{wang2015quantum,wang2018multidimensional,chen2021bright}.

		\begin{figure*}
			\includegraphics[width=1\linewidth]{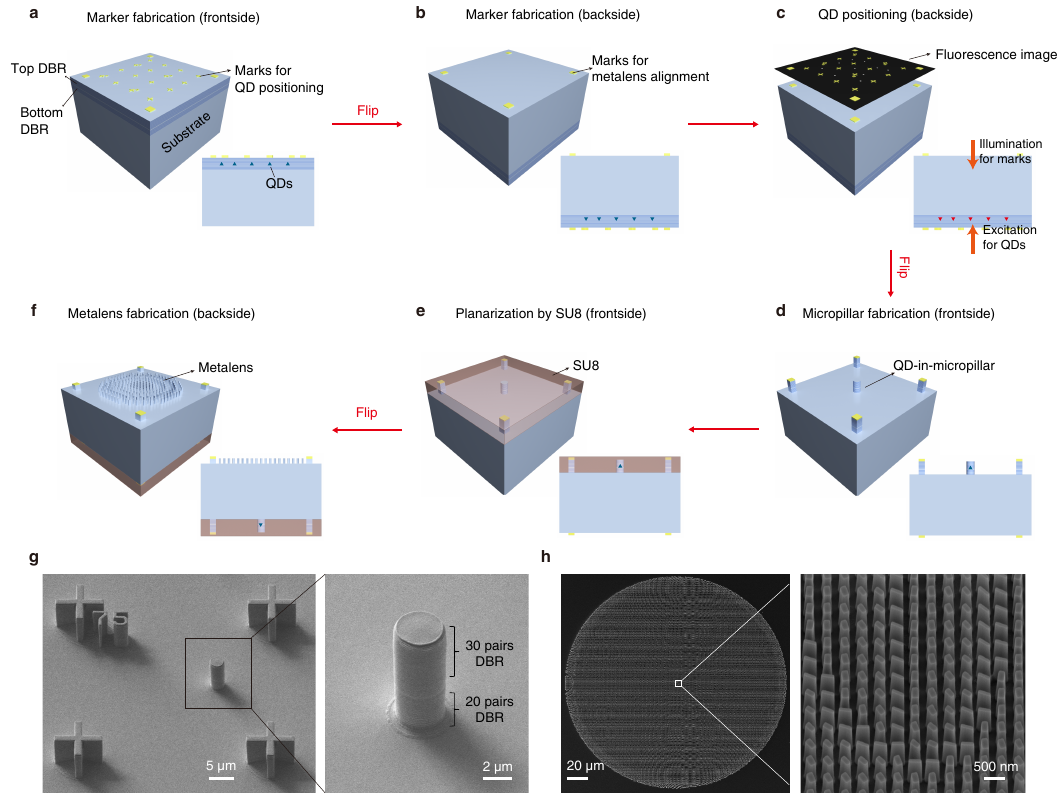}
			\caption{\textbf{Fabrication process of monolithic microcavity-metalens interface.} (a) Fabrication of the front side alignment marks for extracting the positions of QDs. (b) Fabrication of the backside metallic marks to align the metalenses with the front side micropillars. (c) Fluorescence imaging of QDs by illuminating the marks from the back side and exciting the QDs from the front side. The spatial positions of QDs with respect to the front side alignment marks can be extracted from the fluorescence images. (d) Fabrication of the micropillars with QDs in the center by using an aligned EBL and dry etching process. (e) Planarization of front-side micropillars with SU-8 polymer, which protects the micropillars in the fabrication process of metalenses. (f) Fabrication of the backside metalens by another aligned EBL and dry etching process. Insets illustrate the cross-sections in each fabrication step. (g) SEM images of a fabricated front-side micropillar with different magnifications. (h) SEM images of fabricated back-side metalens with different magnifications.}
			\label{fig:Fig2}
		\end{figure*}

		\section{Operation principle of monolithic microcavity-metalens interfaces}
		
		The monolithic interface consists of a pair of a QD-micropillar single-photon source and an ultra-thin metalens, which are accurately aligned on the front side and back side of an III-V compound semiconductor chip, respectively, as schematically shown in Fig.~\ref{fig:Fig1}(a). The micropillar and metalens are decoupled in the near-field, enabling independent optimizations of the single-photon generation and the wavefront shaping. A single epitaxially grown InAs QD is positioned in the center of a 3~$\mu$m-diameter micropillar with 30 (20) pairs of top (bottom) 1/4~$\lambda$ $\rm Al_{0.9}Ga_{0.1}As$/GaAs distributed Bragg reflectors (DBRs), as presented in Fig.~\ref{fig:Fig1}(b). The high-quality (Q) factor and the wavelength-scale mode volume associated with the micropillar allow the enhanced spontaneous emission rate and improved photon extraction efficiency for the single photons emitted from the QD, as numerically simulated in Fig.~\ref{fig:Fig1}(c). Due to the higher reflectivity of the top DBR, the single photons emitted from the QD are guided towards the GaAs substrate which is transparent at the emission wavelength, see Supplementary Materials(SM) and numerical simulation in Fig.~\ref{fig:Fig1}(f). The emitted single photons are in a Gaussian-like mode\cite{somaschi2016near,ding2016demand,wei2022tailoring} with a divergence angle of $8.25^{\circ}$ at the output of the micropillar, as shown in Fig.~\ref{fig:Fig1}(g). By propagating through a 350~$\mu$m-thickness substrate, the beam diameter expands from $\sim$2~$\mu$m to $\sim$103~$\mu$m, see Fig.~\ref{fig:Fig1}(h),  which is crucial to achieve mode matching between the microcavity and the metalens for efficient wavefront shaping. On the backside of the chip, a metalens with a diameter of 200~$\mu$m is aligned to the emitted single photons for arbitrary phase front engineering. The metalens is composed of cuboid-shaped GaAs meta-atoms, whose height (H), lengths in X (L) and Y (W) directions, periodicities ($P_{x}$, $P_{y}$) in X and Y directions, and orientation angle ($\theta$) can be well controlled to achieve a variety of functionalities, as shown in Fig.~\ref{fig:Fig1}(d). Each GaAs meta-atom serves as a truncated rectangular waveguide, being able to introduce desired and independent phase shifts ($\varphi$) to the right-handed circularly polarized (RCP) light and left-handed circularly polarized (LCP) light. The phase from 0 to 2$\pi$ is digitized into 8 pieces by varying L and W with transmissions above 0.6, as shown in Fig.~\ref{fig:Fig1}(e). The transmission of the meta-atom can be improved by using smaller periodicities ($P_{x}$, $P_{y}$), which is however technologically challenging. The orientation angle ($\theta$) provides an additional degree of freedom to engineer the Pancharatnam-Berry (PB) phase\cite{berry1984quantal}, which is crucial for the decoupling and independent phase control of the RCP and LCP channels.

		\begin{figure*}
			\begin{center}
				\includegraphics[width=1\linewidth]{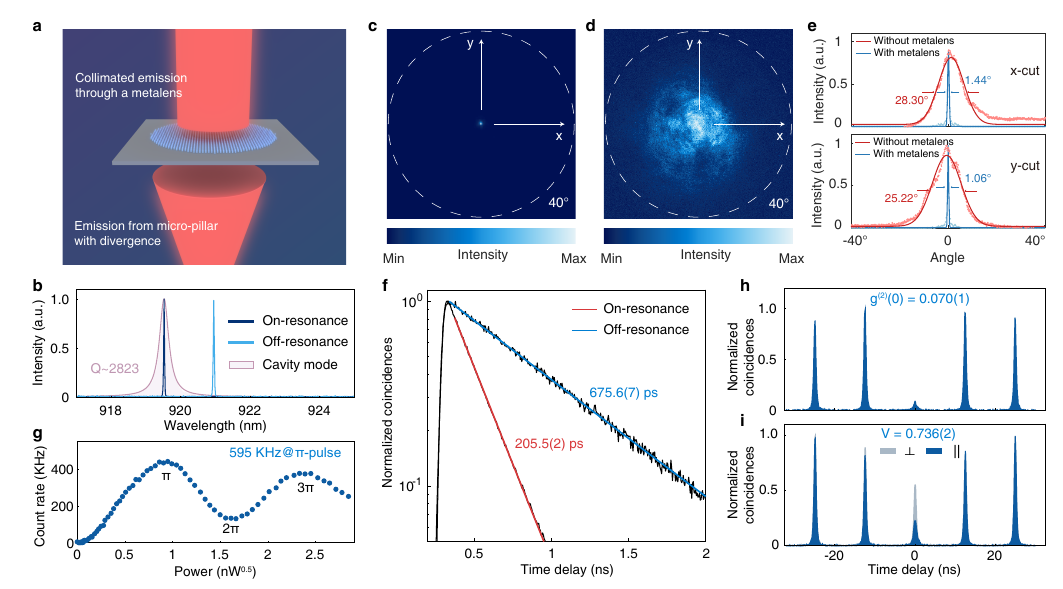}
				\caption{\textbf{Characterization of the metalens-collimated single-photon source.}  (a) Schematics of the metalens-collimated single-photon source. A QD in the micropillar emits single photons to the substrate with a divergent angle of $8.25^{\circ}$. The single-photon light beam is collimated by a metalens on the backside, giving rise to collimated single-photon emission. (b) $\mu$PL spectra for the QD resonant with and detuned from the cavity mode. (c) Far-field pattern of the collimated single-photon emission. (d) Far-field pattern of a reference single-photon emission without the collimation metalens. (e) Comparison of the cross sections of far-field patterns for the single-photon emission with and without the collimation lens. The one with collimation metalens features a divergence angle as small as $1.44^{\circ}$($1.06^{\circ}$) while the single-photon emission without collimation exhibits a large divergence angle of $28.30^{\circ}$($25.22^{\circ}$). (f) Lifetimes of the QD when on- and off-resonance to the cavity mode. A spontaneous emission rate enhancement of 3.3 is extracted. (g) Resonance fluorescence intensity as a function of the power of the pulsed excitation laser. Clear Rabi-oscillation is observed, revealing the coherent nature of a quantum two-level system. (h, i) Coincidence histogram of Hanbury-Brown-Twiss (HBT) and Hong-Ou-Mandel (HOM) interferences for the single photons under $\pi$-pulsed excitation, giving rise to a $g^{(2)}(0)$ = 0.070(1) and $V_{\rm{HOM}}$ = 0.736(2), respectively. The $g^{(2)}(0)$ is calculated from the integrated area in the zero-delay peak divided by the mean of the peaks away from the zero delay and $V_{\rm{HOM}}$ is calculated from the ratio of the central peak area between the co-polarization and cross-polarization. The $g^{(2)}(0)$ value is corrected in the extracted $V_{\rm{HOM}}$.}
				\label{fig:Fig3}
			\end{center}
		\end{figure*}
		
		\section{Fabrication of monolithic microcavity-metalens interfaces}
		
		We have developed a highly customized fabrication process to implement our proposed device, as schematically shown in Fig.~\ref{fig:Fig2}. A planar cavity consisting of 20 pairs of bottom 1/4 $\lambda$ $\rm Al_{0.9}Ga_{0.1}As$/GaAs DBR, a $\lambda$ GaAs cavity containing single InAs QDs and 30 pairs of top 1/4 $\lambda$ $\rm Al_{0.9}Ga_{0.1}As$/GaAs DBR is grown on a 350 $\mu$m double-side polished GaAs substrate via molecule beam epitaxy. Two sets of metallic marks which are used for positioning QDs in the micropillar and aligning metalens to the micropillar are fabricated on the front side of the wafer by using an electron-beam lithography (EBL) and a lift-off process (Fig.~\ref{fig:Fig2}(a)). A double-sided photolithography and a lift-off process are used to create alignment marks on the backside of the chip (Fig.~\ref{fig:Fig2}(b)). The alignment accuracy between the front side marks and back side marks is within 1 $\mu$m (see Methods and SM), ensuring that all the emitted single photons can be modulated by the metalens. The spatial positions of individual QDs respective to the front alignment marks are extracted by using our well-established fluorescence imaging technique (Fig.~\ref{fig:Fig2}(c))\cite{liu2017cryogenic,liu2024super}. Deterministically coupled QD-micropillar single-photon sources are realized by using a second aligned EBL and a chlorine-based dry etching  (Fig.~\ref{fig:Fig2}(d)). The front side of the chip is then planarized with SU-8 photoresist to protect the micropillars in the following metalens fabrication process (Fig.~\ref{fig:Fig2}(e)). Finally, metalenses are fabricated on the back side of the chip via the third aligned EBL and a dry etching process (Fig.~\ref{fig:Fig2}(f)). The scanning electron microscope (SEM) images with different magnifications of a fabricated micropillar and a metalens are presented in Fig.~\ref{fig:Fig2}(g, h). We systematically optimized the EBL exposure parameters and dry etching recipes for micropillars and metalenses individually, leading to high-quality fabrication with vertical etching profiles and excellent surface smoothness.
		
		\begin{center}
			\begin{figure*}
				\begin{center}
					\includegraphics[width=1\linewidth]{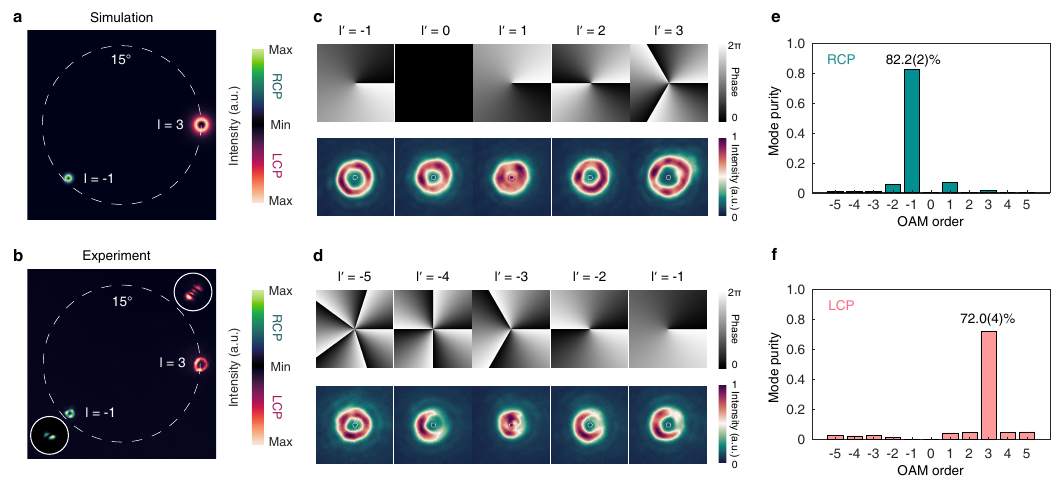}
					\caption{\textbf{Manipulating the photon states of the emitted single photons.} (a) Simulated far-field patterns for the metalens with multi-dimensional functionality. (b) Measured far-field patterns for the metalens with multi-dimensional functionality. Insets in (b) are the interference patterns of the quantum vortices with an offset, clearly revealing the topological charges of the OAM. The topological numbers of the structured single photons are quantitatively extracted with $\rm{L_{RCP}}$ = -1 and $\rm{L_{LCP}}$ = 3. (c,d) Measured far-field profiles of the RCP and LCP channel in (b) by projecting them to a series of vortex wave plates (VWPs) with different topological charge of $l'$, respectively. In the bottom panels of (c) and (d), the white dashed circles indicate the central regions used to extract the OAM-mode spectrum. (e,f) Measured OAM mode spectrum of the RCP and LCP channels in (b), respectively. The OAM purities are 82.2(2)\% for $\rm{L_{RCP}}$ = -1 and 72.0(4)\% for $\rm{L_{LCP}}$ = 3.}
					\label{fig:Fig4}
				\end{center}
			\end{figure*}
		\end{center} 
		
		\begin{center}
			\begin{figure*}
				\begin{center}
					\includegraphics[width=1\linewidth]{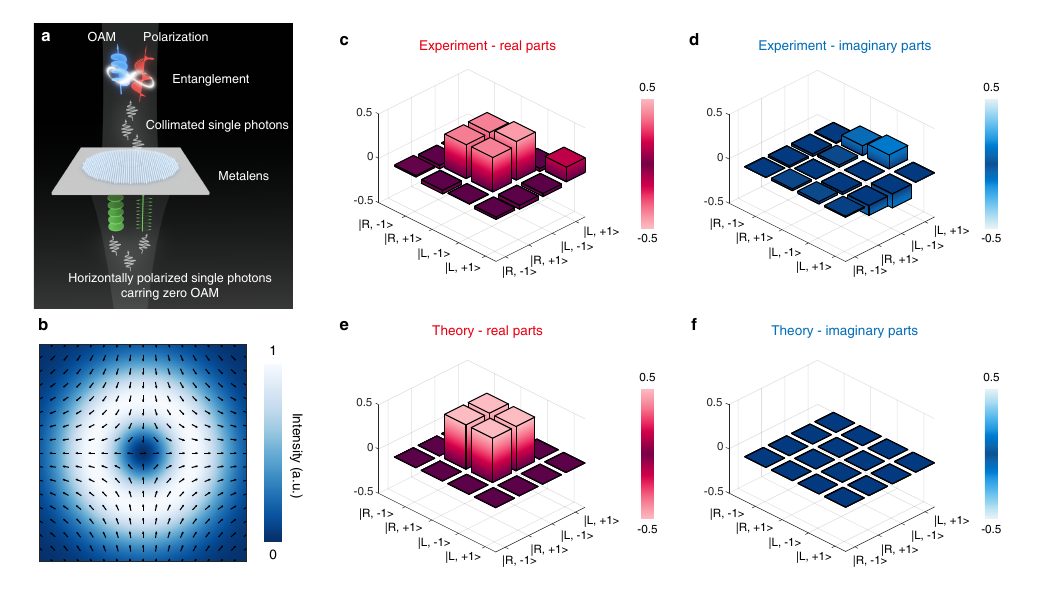}
					\caption{\textbf{Generation of polarization-OAM entanglement with micropillar-metalens interfaces.} (a) Schematic of entanglement between polarization and OAM on a single photon. A horizontally polarized single photon goes into the metalens. This photon carries zero OAM, as illustrated by the green flat phase front. The single photon passes through the metalens and comes out as a single-particle entangled state, depicted as a superposition of the red and blue electric field amplitudes, with the corresponding vortex phase fronts opposite to one another. (b) Intensity and polarization patterns of the Bell state in the combined OAM and polarization space. (c) Real and (d) imaginary parts of the measured density matrix for the state reconstructed via quantum state tomography. The fidelity between the reconstructed state and the theoretical one is equal to $F = 0.8914\pm 0.0023$, where the standard deviation is estimated through a Monte-Carlo analysis assuming a Poissonian statistics. (e) Real and (f) imaginary parts of the density matrix for the ideal polarization-OAM entangled state.}
					\label{fig:Fig5}
				\end{center}
			\end{figure*}
		\end{center}
		
		\begin{center}
			\begin{figure*}
				\begin{center}
					\includegraphics[width=1\linewidth]{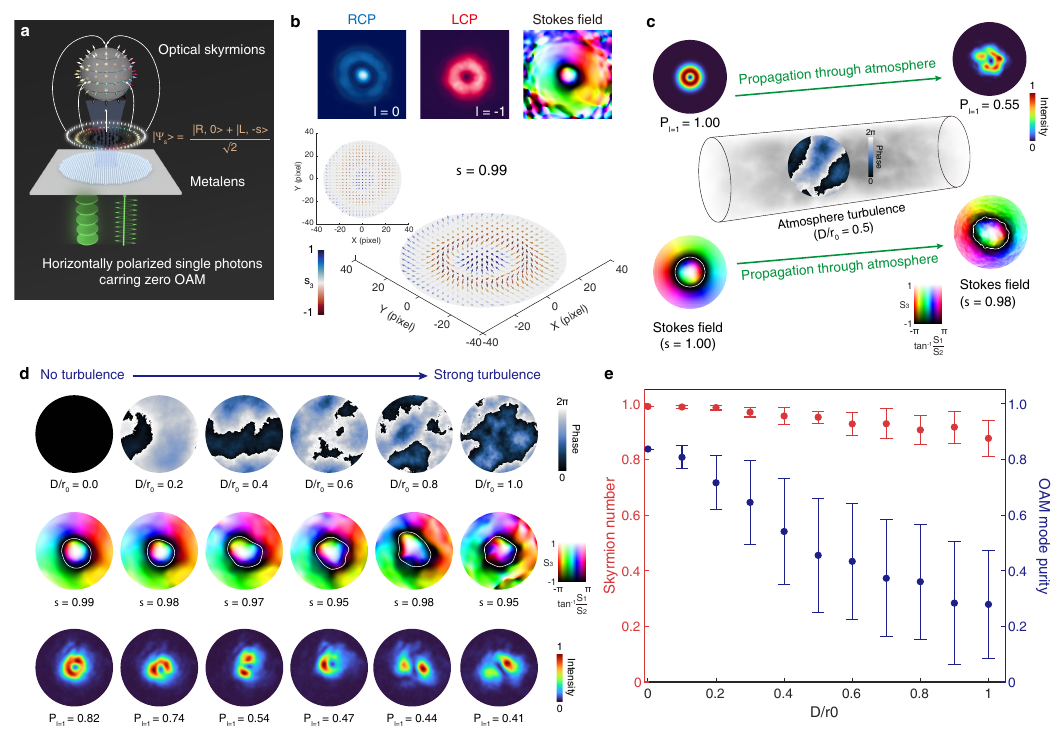}
					\caption{\textbf{Generation of single-photon skyrmions with topological stability by} micropillar-metalens interfaces. (a) Schematic of single-photon skyrmions generated by micropillar-metalens interfaces. (b) The normalized Stokes vector distributions with skyrmion number equal to 1, generated by a simultaneous complex-amplitude-modulation metalens device. The resulting Stokes field exhibits a well-defined skyrmion boundary. Insets in (b) present the measured far-field patterns for the RCP and LCP components and the corresponding Stokes fields. (c) Theoretical model depicting the topological protection of single-photon skyrmions in a turbulent environment. Despite distortions in the Stokes vector field and its skyrmion boundary, the skyrmion number of the Stokes field can still be accurately recovered via boundary identification. (d) Experimentally measured Stokes vector field and the corresponding recovered skyrmion numbers for single-photon skyrmions after propagation through single realizations of atmospheric turbulence with different strengths. (e) Evolution of skyrmion number and OAM mode purity versus turbulence strength $D/r_{0}$, after the single photons with skyrmionic texture ($s=1$) and OAM ($l=1$) propagate through atmospheric turbulence, respectively. For each strength of the turbulence, 50 realizations are used in our measurements. These statistics correspond to the error bar. In (b-d) and throughout this paper, the hue and the lightness of the Stokes field pattern represent the different in-plane and out-of-plane components of the Stokes vectors, respectively. While for the vector arrow distributions, the in-plane and out-of-plane components of the Stokes vectors are determined by the direction and color of the arrows.}
					\label{fig:Fig6}
				\end{center}
			\end{figure*}
		\end{center}
		
		\section{Bright and indistinguishable single-photon emission from micropillar-metalens interfaces}
		
		In the experiment, we first examine the characteristics of the emitted single photons whose emission divergence angle is squeezed by using a collimation metalens, as schematically shown in Fig.~\ref{fig:Fig3}(a). The micro-photoluminescence ($\mu$PL) spectra of the micropillar-metalens system under non-resonant excitation are presented in Fig.~\ref{fig:Fig3}(b). A Q-factor of 2823 is extracted for the investigated micropillar cavity by fitting the high-power cavity spectrum (the broad peak) with a Lorentzian function. Very sharp emission lines (narrow peaks) associated with single QDs are clearly observed at low excitation powers. Detuning between the QDs and the cavity mode can be controlled by varying the system's temperature. The far-field radiation of QD emission after the metalens is measured by using an angle-resolved spectroscopy, exhibiting a beam divergence angle much smaller than that of a reference device without passing any collimation metalens, as shown in Fig.~\ref{fig:Fig3}(c,d). The X-cut and Y-cut of the far-field patterns in Fig.~\ref{fig:Fig3}(c,d) are presented in Fig.~\ref{fig:Fig3}(e), revealing a beam divergence angle of $1.44^{\circ}$($1.06^{\circ}$) and $28.30^{\circ}$($25.22^{\circ}$) in X(Y) direction for the collimated and uncollimated quantum emission, respectively. In Fig.~\ref{fig:Fig3}(f), time-resolved measurements of the QD (a negatively charged exciton) under on-resonance and off-resonance conditions reveal an appreciable reduction of the lifetime from 675.6(7) ps to 205.5(2) ps, which corresponds to a spontaneous emission rate enhancement by a factor of $\sim$3.288(5). The deviation of the measured Purcell factor from the calculated value of 5.37 by using a measured Q-factor of 2830(1) (see details in SM) is probably because of the reduced accuracy in positioning process in which both QDs and the alignment marks are imaged through a 350 $\mu$m substrate. We further perform pulsed resonant excitation on the investigated single QD under the transmission configuration. A clear Rabi oscillation behavior as a function of the excitation power (Fig.~\ref{fig:Fig3}(g)) is observed, which is consistent with the coherent nature of a two-level quantum system. The brightness of the source is evaluated from the measured single-photon count rate of 595 kHz in the avalanche photodiode (APD) at $\pi$-pulse and the calibrated setup transmission, resulting in an extraction efficiency of 35.7(5)\% at the first collection lens. We note that this value already includes the loss induced by the metalens. On the other hand, the single-photon purity and the photon indistinguishability mainly relate to the excitation method and the charge environment of the QDs, which are not expected to decrease with the implementation of the metalens. The coupling efficiency of the emitted single photons into a single-mode fiber is $\sim$0.65 by using the free space path as a reference, which is competitive to the state-of-the-art\cite{rickert2025fiber, margaria2024efficient}. The indistinguishability of the emitted single photons at $\pi$-pulse is measured by using Hanbury-Brown-Twiss and Hong-Ou-Mandel interferometers, giving rise to a multi-photon suppression of a $g^{(2)}(0)$ = 0.070(1) and photon indistinguishability with a visibility of $V_{\rm{HOM}}$ = 0.736(2) respectively, as presented in Fig.~\ref{fig:Fig3}(h,i). The measured indistinguishability is still lower than the state-of-the-art of microcavity-based single-photon sources\cite{somaschi2016near,he2017deterministic,wang2019demand,uppu2020scalable,wei2022tailoring}
		, but serves as the first report for quantum radiation from single quantum emitters integrated with metasurfaces.

		\section{Tailoring quantum emission in multiple degrees of freedom}
		
		After carefully characterizing the collimated single photons emitted from the micropillar-metalens interface, we introduce more complex functionalities to the metalens for the multi-dimensional engineering of quantum emission. We design a representative metalens with distinct functionalities and increasing complexity. This multi-functional device simultaneously focuses and deflects RCP and LCP single photons in different directions, and the single photons at each direction are encoded with different helical phase fronts carrying distinctive OAMs. Considering that the fundamental mode of the micropillar is near-Gaussian\cite{liu2021dual}, the performance of the metalens is numerically simulated by passing a Gaussian beam, which mimics the single photons coupled to the cavity mode, across the designed phase profiles, as presented in Fig.~\ref{fig:Fig4}(a). The experimental verification of the proposed device functionality is presented in Fig.~\ref{fig:Fig4}(b). The RCP and LCP single photons with different values are focused and deflected to the targeted directions, which is in excellent agreement with Fig.~\ref{fig:Fig4}(a). We employ a simplified method to characterize the emitted quantum optical vortices by using offset interference patterns, which was successfully demonstrated in a recent work\cite{li2023arbitrarily}. The topological numbers of the structured single photons are quantitatively extracted with $\rm{L_{RCP}}$ = -1 and $\rm{L_{LCP}}$ = 3 in Fig~\ref{fig:Fig4}(b), which faithfully produces the results in the numerical simulations as presented in Fig.~\ref{fig:Fig4}(a). To further evaluate the purity of the OAM states, we performed projection measurements by passing the emitted single photons from the RCP and LCP channels through a series of vortex wave plates (VWPs) with varying topological charges $l'$, as illustrated in Fig.~\ref{fig:Fig4}(c,d), respectively. The corresponding OAM mode spectra, presented in Fig.~\ref{fig:Fig4}(e,f), were obtained from these projection results. The OAM purities of 82.2(2)\% and 72.0(4)\% are extrated for $\rm{L_{RCP}}$ = -1 and $\rm{L_{LCP}}$ = 3 (see Methods), respectively. As a result, our micropillar-metalens interfaces convincingly demonstrate full and independent control of bright and indistinguishable single photons in radiation divergence, emission directionality, polarization state and OAM dimensions, simultaneously.

		\section{Generation of polarization-OAM entanglement in single photons}
		
		We further demonstrate the generation of quantum entanglement between the polarization state and the OAM of the single photons by using the micropillar-metalens interface. Under LA-phonon-assisted excitation\cite{thomas2021bright,coste2023high}, we can generate a stream of horizontally polarized single photons without OAM, denoted as $\vert$H, 0$\rangle$, as shown in Fig.~\ref{fig:Fig5}(a). These photons pass a metalens acting as a q-plate with q = 1/2, resulting in the formation of vectorial single photons with non-uniform polarization and intensity distributions, as shown in Fig.~\ref{fig:Fig5}(b). In such a state, polarization state and OAM are entangled in the form of Bell state $\vert\Phi_{+}\rangle$ = 1/$\sqrt{2}$ ($\vert$L, -1$\rangle$ +$\vert$R, 1$\rangle$)\cite{stav2018quantum,suprano2023orbital}, as schematically illustrated in Fig.~\ref{fig:Fig5}(a). The maximally entangled Bell state with an entanglement fidelity of $F = 0.8914\pm 0.0023$ is experimentally measured by using quantum state tomography (QST) and reconstructed via a maximum likelihood estimation method\cite{james2001measurement}. The retrieved density matrix is shown in Fig.~\ref{fig:Fig5}(c,d), which is in excellent agreement with the theoretical values in Fig.~\ref{fig:Fig5}(e,f). 
		
		\section{Creation of single photons with local topological spin textures}
		Finally, we present the versatile creations of quantum emission with local topological spin textures, namely single-photon skyrmions, by using our devices. It has been very recently shown that the non-local correlations between polarization and OAM associated with a photon pair can be exploited as a resilient resource of quantum entanglement\cite{ornelas2024non}. On the other hand, the quantum state of light carrying local skyrmionic texture is highly desirable due to the applications in high-capacity quantum communication and high-density quantum memory, yet their on-chip implementation is still elusive\cite{shen2024optical,shen2025free}. Single-photon skyrmions have been very recently generated in a solid-state cavity quantum electrodyamics system under a strong magnetic field\cite{ma2025nanophotonic}; however, such a device is only capable of producing very specific optical skyrmions with a fixed topological charge, without revealing the advantages of such topologically structured light. Optical skyrmions with local topologies can be synthesized via the superposition of two light beams with orthogonal circular polarizations and different OAM orders\cite{shen2022generation, lin2024chip, he2024optical}. As schematically shown in Fig.~\ref{fig:Fig6}(a), the horizontally polarized single photons with zero OAM pass through the metalens, encoding a helical phase front to the LCP component. The created polarization state features a topological texture in the form of an optical skyrmion, which can be written as $\vert\Psi_{s}\rangle = 1/\sqrt{2} (\vert R, 0 \rangle +\vert L, -s \rangle$), where the skyrmion number $s$ equals to the OAM value of the LCP component. We experimentally generate optical skyrmions with skyrmion numbers equal to 1, as shown in Fig.~\ref{fig:Fig6}(b). Within the cross-section of the optical beam, the polarization vector flips upward while swirling, and all the polarization states represented by the unit Poincaré sphere are experienced, thus fulfilling the skyrmion topology. From the measured Stokes vectors, skyrmion numbers of 0.99 are extracted (see Methods), which are very close to theoretical values of 1. To further demonstrate the functionality of our microcavity-metalens interfaces, we generate optical skyrmions with skyrmion numbers ranging from 2 to 5. One of the greatest promises of topologically structured light is to provide robust propagation against disorders\cite{klug2021orbital}, which has recently stimulated great interests in the stability of optical skyrmions\cite{chen2025more,he2022towards,shen2021supertoroidal}. The propagation of single-photon skyrmions in an atmospheric turbulence can be numerically simulated, as shown for one example in Fig.~\ref{fig:Fig6}(c). Despite local changes in the spin textures, the skymion number is slightly reduced from an ideal value of 1.00 down to 0.98 under a turbulence strength with ratio $D/r_{0} = 0.5$ (see Methods). The simulation is experimentally verified by examining the skyrmion numbers of the single-photon skyrmions propagating through artificial atmospheric turbulences with different strengths, which are created by a spatial phase modulator (SLM)\cite{cox2020structured,toselli2015slm}, see more details in Methods. While local variations are observed in the Stokes parameters and the skyrmion boundaries, the skyrmion number remains stable under different strengths of perturbation, as shown in Fig.~\ref{fig:Fig6}(d), suggesting a resilient propagation related to the topology. For comparison, we also test propagation of single photons carrying OAM (RCP channel of Fig.~\ref{fig:Fig4}(b)) through the same atmospheric turbulences. Single photons with skyrmionic textures exhibit greater robustness against atmospheric turbulence than those carrying OAM with the same topological charge. The statistical measurements (represented by the error bars), based on 50 realizations for each strength of the turbulence, are shown in Fig.~\ref{fig:Fig6}(e). The local topologies carried by single-photon skyrmions could potentially provide an unparalleled opportunity to access resilient quantum technology such as robust quantum entanglement generations\cite{wang2018multidimensional}, high-dimensional quantum communications\cite{wang2015quantum,chen2021bright} and high-capacity quantum memories\cite{liu2021heralded}.

		\section{Conclusion}
		
		We have successfully demonstrated monolithic microcavity-metalens interfaces capable of generating indistinguishable quantum emission and simultaneously manipulating their optical states in multiple physical degrees of freedom. The cQED effect associated with the micropillar enables single-photon emission with simultaneous high degrees of source brightness, single-photon purity and photon indistinguishability while the metalenses endow the ability to independently manipulate the quantum emission in radiation divergence, emission directionality, polarization state and OAM dimension. The polarization-OAM entanglement and single-photon skyrmions are successfully created in our monolithic devices. The all-in-one functionalities in a single chip offer unique advantages in terms of device footprint, operation stability, and scalability, placing our device as a record-performance source among the existing high-dimensional single-photon sources based on quantum emitters integrated with metasurfaces. Our devices are particularly appealing to the pursuit of a quantum network with larger information capacity, resilience to noise and improved security\cite{erhard2020advances}. From a view of fundamental physics, the investigation of structured light-matter interactions down to the single quanta level may bring new insight into chiral quantum emission\cite{lodahl2017chiral} and spin-photon quantum interfaces\cite{coste2023high,cogan2023deterministic}. Our work bridges the research field of quantum photonics and meta-optics, opening up an avenue toward high-dimensional multi-photon experiments in which device scalability and system efficiency play an essential role.

		
		\vspace{0.8em}\noindent  \textbf{Data availability}
		
		\noindent{The data that support the findings of this study are available within the paper and the Supplementary Materials. Other relevant data are available from the corresponding authors on reasonable request.}
		
		\vspace{0.8em}\noindent  \textbf{Acknowledgements}
		
		\noindent{This research was supported by the National Key Research and Development Program of China (2021YFA1400800); National Natural Science Foundation of China (62035017, 12361141824, 12304409); Natural Science Foundation of Guangdong Province (2023B1515120070); Guangdong Provincial Quantum Science Strategic Initiative (GDZX2206001, GDZX2306003); Singapore Ministry of Education (MOE) AcRF Tier 1 grants (RG157/23 \& RT11/23); Singapore Agency for Science, Technology and Research (A*STAR) MTC Individual Research Grants (M24N7c0080) and the National Super-Computer Center in Guangzhou.}
		
		\vspace{0.8em}\noindent  \textbf{Author Contributions}
		
		\noindent J.~L. conceived the project; S.~F.~L., J.~L., and H.~Q.~L. designed the epitaxial structure; H.~Q.~L. and H.~Q.~N. grew the quantum dot wafers; J.~T.~M., S.~F.~L., B.~C., G.~X.~Q. N.~M-C, and D.~Z.~W.developed the theory model and designed the devices; D.~L., J.~T.~M., S.~F.~L., J.~W.~Y., K.~X.~C., and X.~S.~L. fabricated the devices; J.~T.~M. and S.~F.~L. built the setup and performed the optical measurements;  J.~T.~M., Y.~J.~S. and J.~L. analyzed the data; J.~L. and J.~T.~M. prepared the manuscript with inputs from all authors, Z.~C.~N. Y.~Y., Y.~J.~S., L.~L., X.~H.~W. and J.~L. supervised the project.
		
		\vspace{0.8em}\noindent  \textbf{Conflict of Interest}
		
		\noindent The authors declare no conflict of interest.
		
		\bibliographystyle{naturemag}
		\bibliography{bibfile-new}

@article{o2009photonic,
	title={Photonic quantum technologies},
	author={O'Brien, Jeremy L and Furusawa, Akira and Vu{\v{c}}kovi{\'c}, Jelena},
	journal={Nat. Photon.},
	volume={3},
	number={12},
	pages={687--695},
	year={2009},
	publisher={Nature Publishing Group UK London}
}

@article{ma2025nanophotonic,
  title={Nanophotonic quantum skyrmions enabled by semiconductor cavity quantum electrodynamics},
  author={Ma, Jiantao and Yang, Jiawei and Liu, Shunfa and Chen, Bo and Li, Xueshi and Song, Changkun and Qiu, Guixin and Zou, Kai and Hu, Xiaolong and Li, Feng and others},
  journal={Nature Physics},
  volume={21},
  number={9},
  pages={1462--1468},
  year={2025},
  publisher={Nature Publishing Group UK London}
}

@article{klug2021orbital,
  title={The orbital angular momentum of a turbulent atmosphere and its impact on propagating structured light fields},
  author={Klug, Asher and Nape, Isaac and Forbes, Andrew},
  journal={New Journal of Physics},
  volume={23},
  number={9},
  pages={093012},
  year={2021},
  publisher={IOP Publishing}
}

@article{chen2025more,
  title={More than just a name? From magnetic to optical skyrmions and the topology of light},
  author={Chen, Jian and Forbes, Andrew and Qiu, Cheng-Wei},
  journal={Light: Science \& Applications},
  volume={14},
  number={1},
  pages={28},
  year={2025},
  publisher={Nature Publishing Group UK London}
}

@article{shen2021supertoroidal,
  title={Supertoroidal light pulses as electromagnetic skyrmions propagating in free space},
  author={Shen, Yijie and Hou, Yaonan and Papasimakis, Nikitas and Zheludev, Nikolay I},
  journal={Nature communications},
  volume={12},
  number={1},
  pages={5891},
  year={2021},
  publisher={Nature Publishing Group UK London}
}

@article{he2022towards,
  title={Towards higher-dimensional structured light},
  author={He, Chao and Shen, Yijie and Forbes, Andrew},
  journal={Light: Science \& Applications},
  volume={11},
  number={1},
  pages={205},
  year={2022},
  publisher={Nature Publishing Group UK London}
}

@article{cox2020structured,
  title={Structured light in turbulence},
  author={Cox, Mitchell A and Mphuthi, Nokwazi and Nape, Isaac and Mashaba, Nikiwe and Cheng, Ling and Forbes, Andrew},
  journal={IEEE Journal of Selected Topics in Quantum Electronics},
  volume={27},
  number={2},
  pages={1--21},
  year={2020},
  publisher={IEEE}
}

@article{toselli2015slm,
  title={SLM-based laboratory simulations of Kolmogorov and non-Kolmogorov anisotropic turbulence},
  author={Toselli, Italo and Korotkova, Olga and Xiao, Xifeng and Voelz, David G},
  journal={Applied optics},
  volume={54},
  number={15},
  pages={4740--4744},
  year={2015},
  publisher={Optical Society of America}
}

@article{he2024optical,
  title={Optical skyrmions from metafibers with subwavelength features},
  author={He, Tiantian and Meng, Yuan and Wang, Lele and Zhong, Hongkun and Mata-Cervera, Nilo and Li, Dan and Yan, Ping and Liu, Qiang and Shen, Yijie and Xiao, Qirong},
  journal={Nature Communications},
  volume={15},
  number={1},
  pages={10141},
  year={2024},
  publisher={Nature Publishing Group UK London}
}

@article{shen2025free,
  title={Free-space topological optical textures: tutorial},
  author={Shen, Yijie and Wang, Haiwen and Fan, Shanhui},
  journal={Advances in Optics and Photonics},
  volume={17},
  number={2},
  pages={295--374},
  year={2025},
  publisher={Optica Publishing Group}
}

@article{shen2024optical,
  title={Optical skyrmions and other topological quasiparticles of light},
  author={Shen, Yijie and Zhang, Qiang and Shi, Peng and Du, Luping and Yuan, Xiaocong and Zayats, Anatoly V},
  journal={Nature Photonics},
  volume={18},
  number={1},
  pages={15--25},
  year={2024},
  publisher={Nature Publishing Group UK London}
}

@article{ornelas2024non,
  title={Non-local skyrmions as topologically resilient quantum entangled states of light},
  author={Ornelas, Pedro and Nape, Isaac and de Mello Koch, Robert and Forbes, Andrew},
  journal={Nature Photonics},
  volume={18},
  number={3},
  pages={258--266},
  year={2024},
  publisher={Nature Publishing Group UK London}
}

@article{james2001measurement,
  title={Measurement of qubits},
  author={James, Daniel FV and Kwiat, Paul G and Munro, William J and White, Andrew G},
  journal={Physical Review A},
  volume={64},
  number={5},
  pages={052312},
  year={2001},
  publisher={APS}
}

@article{kan2023advances,
  title={Advances in metaphotonics empowered single photon emission},
  author={Kan, Yinhui and Bozhevolnyi, Sergey I},
  journal={Advanced Optical Materials},
  volume={11},
  number={10},
  pages={2202759},
  year={2023},
  publisher={Wiley Online Library}
}

@article{jia2023multichannel,
  title={Multichannel Single-Photon Emissions with On-Demand Momentums by Using Anisotropic Quantum Metasurfaces},
  author={Jia, Shangtong and Li, Yongkang and Xue, Zeyang and Chen, Kangyao and Li, Zhi and Gong, Qihuang and Chen, Jianjun},
  journal={Advanced Materials},
  volume={35},
  number={26},
  pages={2212244},
  year={2023},
  publisher={Wiley Online Library}
}

@article{shen2022generation,
  title={Generation of optical skyrmions with tunable topological textures},
  author={Shen, Yijie and Mart{\'\i}nez, Eduardo Casas and Rosales-Guzm{\'a}n, Carmelo},
  journal={ACS Photonics},
  volume={9},
  number={1},
  pages={296--303},
  year={2022},
  publisher={ACS Publications}
}

@article{lin2024chip,
  title={On-chip optical skyrmionic beam generators},
  author={Lin, Wenbo and Ota, Yasutomo and Arakawa, Yasuhiko and Iwamoto, Satoshi},
  journal={Optica},
  volume={11},
  number={11},
  pages={1588--1594},
  year={2024},
  publisher={Optica Publishing Group}
}

@article{liu2021heralded,
  title={Heralded entanglement distribution between two absorptive quantum memories},
  author={Liu, Xiao and Hu, Jun and Li, Zong-Feng and Li, Xue and Li, Pei-Yun and Liang, Peng-Jun and Zhou, Zong-Quan and Li, Chuan-Feng and Guo, Guang-Can},
  journal={Nature},
  volume={594},
  number={7861},
  pages={41--45},
  year={2021},
  publisher={Nature Publishing Group UK London}
}

@article{margaria2024efficient,
  title={Efficient fiber-pigtailed source of indistinguishable single photons},
  author={Margaria, Nico and Pastier, Florian and Bennour, Thinhinane and Billard, Marie and Ivanov, Edouard and Hease, William and Stepanov, Petr and Adiyatullin, Albert F and Singla, Raksha and Pont, Mathias and others},
  journal={arXiv preprint arXiv:2410.07760},
  year={2024}
}

@article{rickert2025fiber,
  title={A fiber-pigtailed quantum dot device generating indistinguishable photons at GHz clock-rates},
  author={Rickert, Lucas and {\.Z}o{\l}nacz, Kinga and Vajner, Daniel A and von Helversen, Martin and Rodt, Sven and Reitzenstein, Stephan and Liu, Hanqing and Li, Shulun and Ni, Haiqiao and Wyborski, Pawe{\l} and others},
  journal={Nanophotonics},
  number={0},
  year={2025},
  publisher={De Gruyter}
}

@article{chen2020flat,
  title={Flat optics with dispersion-engineered metasurfaces},
  author={Chen, Wei Ting and Zhu, Alexander Y and Capasso, Federico},
  journal={Nat. Rev. Mater.},
  volume={5},
  number={8},
  pages={604--620},
  year={2020},
  publisher={Nature Publishing Group UK London}
}

@article{chen2016review,
  title={A review of metasurfaces: physics and applications},
  author={Chen, Hou-Tong and Taylor, Antoinette J and Yu, Nanfang},
  journal={Rep. Prog. Phys.},
  volume={79},
  number={7},
  pages={076401},
  year={2016},
  publisher={IOP Publishing}
}

@article{elshaari2020hybrid,
  title={Hybrid integrated quantum photonic circuits},
  author={Elshaari, Ali W and Pernice, Wolfram and Srinivasan, Kartik and Benson, Oliver and Zwiller, Val},
  journal={Nat. Photon.},
  volume={14},
  number={5},
  pages={285--298},
  year={2020},
  publisher={Nature Publishing Group UK London}
}

@article{solntsev2021metasurfaces,
  title={Metasurfaces for quantum photonics},
  author={Solntsev, Alexander S and Agarwal, Girish S and Kivshar, Yuri S},
  journal={Nat. Photon.},
  volume={15},
  number={5},
  pages={327--336},
  year={2021},
  publisher={Nature Publishing Group UK London}
}

@article{ding2023advances,
  title={Advances in quantum meta-optics},
  author={Ding, Fei and Bozhevolnyi, Sergey I},
  journal={Mater. Today},
  year={2023},
  publisher={Elsevier}
}

@article{wang2018multidimensional,
  title={Multidimensional quantum entanglement with large-scale integrated optics},
  author={Wang, Jianwei and Paesani, Stefano and Ding, Yunhong and Santagati, Raffaele and Skrzypczyk, Paul and Salavrakos, Alexia and Tura, Jordi and Augusiak, Remigiusz and Man{\v{c}}inska, Laura and Bacco, Davide and others},
  journal={Science},
  volume={360},
  number={6386},
  pages={285--291},
  year={2018},
  publisher={American Association for the Advancement of Science}
}

@article{suprano2023orbital,
  title={Orbital angular momentum based intra-and interparticle entangled states generated via a quantum dot source},
  author={Suprano, Alessia and Zia, Danilo and Pont, Mathias and Giordani, Taira and Rodari, Giovanni and Valeri, Mauro and Piccirillo, Bruno and Carvacho, Gonzalo and Spagnolo, Nicol{\`o} and Senellart, Pascale and others},
  journal={Adv. Photon.},
  volume={5},
  number={4},
  pages={046008--046008},
  year={2023},
  publisher={Society of Photo-Optical Instrumentation Engineers}
}

@article{liu2024super,
  title={Super-resolved snapshot hyperspectral imaging of solid-state quantum emitters for high-throughput integrated quantum technologies},
  author={Liu, Shunfa and Li, Xueshi and Liu, Hanqing and Qiu, Guixin and Ma, Jiantao and Nie, Liang and Meng, Yun and Hu, Xiaolong and Ni, Haiqiao and Niu, Zhichuan and others},
  journal={Nat. Photon.},
  volume={18},
  pages={967–974},
  year={2024},
  publisher={Nature Publishing Group UK London}
}

@article{berry1984quantal,
  title={Quantal phase factors accompanying adiabatic changes},
  author={Berry, Michael Victor},
  journal={Proc. R. Soc. London. Ser A},
  volume={392},
  number={1802},
  pages={45--57},
  year={1984},
  publisher={The Royal Society London}
}

@article{yang2024tunable,
  title={Tunable quantum dots in monolithic Fabry-Perot microcavities for high-performance single-photon sources},
  author={Yang, Jiawei and Chen, Yan and Rao, Zhixuan and Zheng, Ziyang and Song, Changkun and Chen, Yujie and Xiong, Kaili and Chen, Pingxing and Zhang, Chaofan and Wu, Wei and others},
  journal={Light Sci. Appl.},
  volume={13},
  number={1},
  pages={33},
  year={2024},
  publisher={Nature Publishing Group UK London}
}

@article{gazzano2013bright,
  title={Bright solid-state sources of indistinguishable single photons},
  author={Gazzano, O and Michaelis de Vasconcellos, S and Arnold, C and Nowak, A and Galopin, E and Sagnes, I and Lanco, L and Lema{\^\i}tre, A and Senellart, P},
  journal={Nat. Commun.},
  volume={4},
  number={1},
  pages={1425},
  year={2013},
  publisher={Nature Publishing Group UK London}
}

@article{wang2018quantum,
  title={Quantum metasurface for multiphoton interference and state reconstruction},
  author={Wang, Kai and Titchener, James G and Kruk, Sergey S and Xu, Lei and Chung, Hung-Pin and Parry, Matthew and Kravchenko, Ivan I and Chen, Yen-Hung and Solntsev, Alexander S and Kivshar, Yuri S and others},
  journal={Science},
  volume={361},
  number={6407},
  pages={1104--1108},
  year={2018},
  publisher={American Association for the Advancement of Science}
}

@article{mckeever2004deterministic,
  title={Deterministic generation of single photons from one atom trapped in a cavity},
  author={McKeever, J and Boca, A and Boozer, AD and Miller, R and Buck, JR and Kuzmich, A and Kimble, HJ},
  journal={Science},
  volume={303},
  number={5666},
  pages={1992--1994},
  year={2004},
  publisher={American Association for the Advancement of Science}
}

@article{giovannetti2011advances,
  title={Advances in quantum metrology},
  author={Giovannetti, Vittorio and Lloyd, Seth and Maccone, Lorenzo},
  journal={Nat. Photon.},
  volume={5},
  number={4},
  pages={222--229},
  year={2011},
  publisher={Nature Publishing Group UK London}
}

@article{o2007optical,
  title={Optical quantum computing},
  author={O'Brien, Jeremy L},
  journal={Science},
  volume={318},
  number={5856},
  pages={1567--1570},
  year={2007},
  publisher={American Association for the Advancement of Science}
}

@article{wang2015quantum,
  title={Quantum teleportation of multiple degrees of freedom of a single photon},
  author={Wang, Xi-Lin and Cai, Xin-Dong and Su, Zu-En and Chen, Ming-Cheng and Wu, Dian and Li, Li and Liu, Nai-Le and Lu, Chao-Yang and Pan, Jian-Wei},
  journal={Nature},
  volume={518},
  number={7540},
  pages={516--519},
  year={2015},
  publisher={Nature Publishing Group UK London}
}

@article{li2023arbitrarily,
  title={Arbitrarily structured quantum emission with a multifunctional metalens},
  author={Li, Chi and Jang, Jaehyuck and Badloe, Trevon and Yang, Tieshan and Kim, Joohoon and Kim, Jaekyung and Nguyen, Minh and Maier, Stefan A and Rho, Junsuk and Ren, Haoran and others},
  journal={Elight},
  volume={3},
  number={1},
  pages={19},
  year={2023},
  publisher={Springer}
}

@article{stav2018quantum,
  title={Quantum entanglement of the spin and orbital angular momentum of photons using metamaterials},
  author={Stav, Tomer and Faerman, Arkady and Maguid, Elhanan and Oren, Dikla and Kleiner, Vladimir and Hasman, Erez and Segev, Mordechai},
  journal={Science},
  volume={361},
  number={6407},
  pages={1101--1104},
  year={2018},
  publisher={American Association for the Advancement of Science}
}

@article{bao2020demand,
  title={On-demand spin-state manipulation of single-photon emission from quantum dot integrated with metasurface},
  author={Bao, Yanjun and Lin, Qiaoling and Su, Rongbin and Zhou, Zhang-Kai and Song, Jindong and Li, Juntao and Wang, Xue-Hua},
  journal={Sci. Adv.},
  volume={6},
  number={31},
  pages={eaba8761},
  year={2020},
  publisher={American Association for the Advancement of Science}
}

@article{li2020metalens,
  title={Metalens-array--based high-dimensional and multiphoton quantum source},
  author={Li, Lin and Liu, Zexuan and Ren, Xifeng and Wang, Shuming and Su, Vin-Cent and Chen, Mu-Ku and Chu, Cheng Hung and Kuo, Hsin Yu and Liu, Biheng and Zang, Wenbo and others},
  journal={Science},
  volume={368},
  number={6498},
  pages={1487--1490},
  year={2020},
  publisher={American Association for the Advancement of Science}
}

@article{aharonovich2016solid,
	title={Solid-state single-photon emitters},
	author={Aharonovich, Igor and Englund, Dirk and Toth, Milos},
	journal={Nat. Photon.},
	volume={10},
	number={10},
	pages={631--641},
	year={2016},
	publisher={Nature Publishing Group UK London}
}

@article{wang2020integrated,
	title={Integrated photonic quantum technologies},
	author={Wang, Jianwei and Sciarrino, Fabio and Laing, Anthony and Thompson, Mark G},
	journal={Nat. Photon.},
	volume={14},
	number={5},
	pages={273--284},
	year={2020},
	publisher={Nature Publishing Group UK London}
}

@article{zhou2023epitaxial,
	title={Epitaxial quantum dots: a semiconductor launchpad for photonic quantum technologies},
	author={Zhou, Xiaoyan and Zhai, Liang and Liu, Jin},
	journal={Photon. Insights},
	volume={1},
	number={2},
	pages={R07--R07},
	year={2023},
	publisher={Society of Photo-Optical Instrumentation Engineers}
}

@article{somaschi2016near,
	title={Near-optimal single-photon sources in the solid state},
	author={Somaschi, Niccolo and Giesz, Valerian and De Santis, Lorenzo and Loredo, JC and Almeida, Marcelo P and Hornecker, Gaston and Portalupi, S Luca and Grange, Thomas and Anton, Carlos and Demory, Justin and others},
	journal={Nat. Photon.},
	volume={10},
	number={5},
	pages={340--345},
	year={2016},
	publisher={Nature Publishing Group UK London}
}

@article{wang2019towards,
	title={Towards optimal single-photon sources from polarized microcavities},
	author={Wang, Hui and He, Yu-Ming and Chung, T-H and Hu, Hai and Yu, Ying and Chen, Si and Ding, Xing and Chen, M-C and Qin, Jian and Yang, Xiaoxia and others},
	journal={Nat. Photon.},
	volume={13},
	number={11},
	pages={770--775},
	year={2019},
	publisher={Nature Publishing Group UK London}
}

@article{he2017deterministic,
	title={Deterministic implementation of a bright, on-demand single-photon source with near-unity indistinguishability via quantum dot imaging},
	author={He, Yu-Ming and Liu, Jin and Maier, Sebastian and Emmerling, Monika and Gerhardt, Stefan and Davan{\c{c}}o, Marcelo and Srinivasan, Kartik and Schneider, Christian and H{\"o}fling, Sven},
	journal={Optica},
	volume={4},
	number={7},
	pages={802--808},
	year={2017},
	publisher={Optica Publishing Group}
}

@article{uppu2020scalable,
	title={Scalable integrated single-photon source},
	author={Uppu, Ravitej and Pedersen, Freja T and Wang, Ying and Olesen, Cecilie T and Papon, Camille and Zhou, Xiaoyan and Midolo, Leonardo and Scholz, Sven and Wieck, Andreas D and Ludwig, Arne and others},
	journal={Sci. Adv.},
	volume={6},
	number={50},
	pages={eabc8268},
	year={2020},
	publisher={American Association for the Advancement of Science}
}

@article{liu2019solid,
	title={A solid-state source of strongly entangled photon pairs with high brightness and indistinguishability},
	author={Liu, Jin and Su, Rongbin and Wei, Yuming and Yao, Beimeng and Silva, Saimon Filipe Covre da and Yu, Ying and Iles-Smith, Jake and Srinivasan, Kartik and Rastelli, Armando and Li, Juntao and others},
	journal={Nat. Nanotechnol.},
	volume={14},
	number={6},
	pages={586--593},
	year={2019},
	publisher={Nature Publishing Group UK London}
}

@article{wang2019demand,
	title={On-demand semiconductor source of entangled photons which simultaneously has high fidelity, efficiency, and indistinguishability},
	author={Wang, Hui and Hu, Hai and Chung, T-H and Qin, Jian and Yang, Xiaoxia and Li, J-P and Liu, R-Z and Zhong, H-S and He, Y-M and Ding, Xing and others},
	journal={Phys. Rev. Lett.},
	volume={122},
	number={11},
	pages={113602},
	year={2019},
	publisher={APS}
}

@article{tomm2021bright,
	title={A bright and fast source of coherent single photons},
	author={Tomm, Natasha and Javadi, Alisa and Antoniadis, Nadia Olympia and Najer, Daniel and L{\"o}bl, Matthias Christian and Korsch, Alexander Rolf and Schott, R{\"u}diger and Valentin, Sascha Ren{\'e} and Wieck, Andreas Dirk and Ludwig, Arne and others},
	journal={Nat. Nanotechnol.},
	volume={16},
	number={4},
	pages={399--403},
	year={2021},
	publisher={Nature Publishing Group UK London}
}

@article{liu2017cryogenic,
	title={Cryogenic photoluminescence imaging system for nanoscale positioning of single quantum emitters},
	author={Liu, Jin and Davan{\c{c}}o, Marcelo I and Sapienza, Luca and Konthasinghe, Kumarasiri and De Miranda Cardoso, Jos{\'e} Vin{\'\i}cius and Song, Jin Dong and Badolato, Antonio and Srinivasan, Kartik},
	journal={Rev. Sci. Instrum.},
	volume={88},
	number={2},
	pages = {023116},
	year={2017},
	publisher={AIP Publishing}
}

@article{cogan2023deterministic,
	title={Deterministic generation of indistinguishable photons in a cluster state},
	author={Cogan, Dan and Su, Zu-En and Kenneth, Oded and Gershoni, David},
	journal={Nat. Photon.},
	volume={17},
	number={4},
	pages={324--329},
	year={2023},
	publisher={Nature Publishing Group UK London}
}

@article{ding2016demand,
	title={On-demand single photons with high extraction efficiency and near-unity indistinguishability from a resonantly driven quantum dot in a micropillar},
	author={Ding, Xing and He, Yu and Duan, Z-C and Gregersen, Niels and Chen, M-C and Unsleber, S and Maier, Sebastian and Schneider, Christian and Kamp, Martin and H{\"o}fling, Sven and others},
	journal={Phys. Rev. Lett.},
	volume={116},
	number={2},
	pages={020401},
	year={2016},
	publisher={APS}
}

@article{liu2021dual,
	title={Dual-resonance enhanced quantum light-matter interactions in deterministically coupled quantum-dot-micropillars},
	author={Liu, Shunfa and Wei, Yuming and Li, Xueshi and Yu, Ying and Liu, Jin and Yu, Siyuan and Wang, Xuehua},
	journal={Light Sci. Appl.},
	volume={10},
	number={1},
	pages={158},
	year={2021},
	publisher={Nature Publishing Group UK London}
}

@article{wei2022tailoring,
	title={Tailoring solid-state single-photon sources with stimulated emissions},
	author={Wei, Yuming and Liu, Shunfa and Li, Xueshi and Yu, Ying and Su, Xiangbin and Li, Shulun and Shang, Xiangjun and Liu, Hanqing and Hao, Huiming and Ni, Haiqiao and others},
	journal={Nat. Nanotechnol.},
	volume={17},
	number={5},
	pages={470--476},
	year={2022},
	publisher={Nature Publishing Group UK London}
}

@article{coste2023high,
	title={High-rate entanglement between a semiconductor spin and indistinguishable photons},
	author={Coste, N and Fioretto, DA and Belabas, N and Wein, SC and Hilaire, P and Frantzeskakis, R and Gundin, M and Goes, B and Somaschi, N and Morassi, M and others},
	journal={Nat. Photon.},
	volume = {17},
	pages={582--587},
	year={2023},
	publisher={Nature Publishing Group UK London}
}

@article{gisin2007quantum,
  title={Quantum communication},
  author={Gisin, Nicolas and Thew, Rob},
  journal={Nat. photon.},
  volume={1},
  number={3},
  pages={165--171},
  year={2007},
  publisher={Nature Publishing Group UK London}
}

@article{zhong2020quantum,
  title={Quantum computational advantage using photons},
  author={Zhong, Han-Sen and Wang, Hui and Deng, Yu-Hao and Chen, Ming-Cheng and Peng, Li-Chao and Luo, Yi-Han and Qin, Jian and Wu, Dian and Ding, Xing and Hu, Yi and others},
  journal={Science},
  volume={370},
  number={6523},
  pages={1460--1463},
  year={2020},
  publisher={American Association for the Advancement of Science}
}

@article{madsen2022quantum,
  title={Quantum computational advantage with a programmable photonic processor},
  author={Madsen, Lars S and Laudenbach, Fabian and Askarani, Mohsen Falamarzi and Rortais, Fabien and Vincent, Trevor and Bulmer, Jacob FF and Miatto, Filippo M and Neuhaus, Leonhard and Helt, Lukas G and Collins, Matthew J and others},
  journal={Nature},
  volume={606},
  number={7912},
  pages={75--81},
  year={2022},
  publisher={Nature Publishing Group UK London}
}

@article{keller2004continuous,
  title={Continuous generation of single photons with controlled waveform in an ion-trap cavity system},
  author={Keller, Matthias and Lange, Birgit and Hayasaka, Kazuhiro and Lange, Wolfgang and Walther, Herbert},
  journal={Nature},
  volume={431},
  number={7012},
  pages={1075--1078},
  year={2004},
  publisher={Nature Publishing Group UK London}
}

@book{michler2017quantum,
  title={Quantum dots for quantum information technologies},
  author={Michler, Peter},
  volume={237},
  year={2017},
  publisher={Springer}
}

@article{senellart2017high,
  title={High-performance semiconductor quantum-dot single-photon sources},
  author={Senellart, Pascale and Solomon, Glenn and White, Andrew},
  journal={Nat. Nanotechnol.},
  volume={12},
  number={11},
  pages={1026--1039},
  year={2017},
  publisher={Nature Publishing Group}
}

@article{montblanch2023layered,
  title={Layered materials as a platform for quantum technologies},
  author={Montblanch, Alejandro R-P and Barbone, Matteo and Aharonovich, Igor and Atat{\"u}re, Mete and Ferrari, Andrea C},
  journal={Nat. Nanotechnol.},
  volume={18},
  number={6},
  pages={555--571},
  year={2023},
  publisher={Nature Publishing Group UK London}
}

@article{zhou2018room,
  title={Room temperature solid-state quantum emitters in the telecom range},
  author={Zhou, Yu and Wang, Ziyu and Rasmita, Abdullah and Kim, Sejeong and Berhane, Amanuel and Bodrog, Zolt{\'a}n and Adamo, Giorgio and Gali, Adam and Aharonovich, Igor and Gao, Wei-bo},
  journal={Sci. Adv.},
  volume={4},
  number={3},
  pages={eaar3580},
  year={2018},
  publisher={American Association for the Advancement of Science}
}

@article{castelletto2014silicon,
  title={A silicon carbide room-temperature single-photon source},
  author={Castelletto, Stefania and Johnson, BC and Iv{\'a}dy, Viktor and Stavrias, N and Umeda, T and Gali, A and Ohshima, T},
  journal={Nat. Mater.},
  volume={13},
  number={2},
  pages={151--156},
  year={2014},
  publisher={Nature Publishing Group UK London}
}

@article{wu2022room,
  title={Room-temperature on-chip orbital angular momentum single-photon sources},
  author={Wu, Cuo and Kumar, Shailesh and Kan, Yinhui and Komisar, Danylo and Wang, Zhiming and Bozhevolnyi, Sergey I and Ding, Fei},
  journal={Sci. Adv.},
  volume={8},
  number={2},
  pages={eabk3075},
  year={2022},
  publisher={American Association for the Advancement of Science}
}

@article{komisar2023multiple,
  title={Multiple channelling single-photon emission with scattering holography designed metasurfaces},
  author={Komisar, Danylo and Kumar, Shailesh and Kan, Yinhui and Meng, Chao and Kulikova, Liudmila F and Davydov, Valery A and Agafonov, Viatcheslav N and Bozhevolnyi, Sergey I},
  journal={Nat. Commun.},
  volume={14},
  number={1},
  pages={6253},
  year={2023},
  publisher={Nature Publishing Group UK London}
}

@article{huang2019monolithic,
  title={A monolithic immersion metalens for imaging solid-state quantum emitters},
  author={Huang, Tzu-Yung and Grote, Richard R and Mann, Sander A and Hopper, David A and Exarhos, Annemarie L and Lopez, Gerald G and Klein, Amelia R and Garnett, Erik C and Bassett, Lee C},
  journal={Nat. Commun.},
  volume={10},
  number={1},
  pages={2392},
  year={2019},
  publisher={Nature Publishing Group UK London}
}

@article{ma2024engineering,
  title={Engineering Quantum Light Sources with Flat Optics},
  author={Ma, Jinyong and Zhang, Jihua and Horder, Jake and Sukhorukov, Andrey A and Toth, Milos and Neshev, Dragomir N and Aharonovich, Igor},
  journal={Adv. Mater.},
  volume={36},
  number={23},
  pages={2313589},
  year={2024},
  publisher={Wiley Online Library}
}

@article{thomas2021bright,
  title={Bright polarized single-photon source based on a linear dipole},
  author={Thomas, SE and Billard, M and Coste, N and Wein, SC and Priya and Ollivier, H and Krebs, Olivier and Taza{\"\i}rt, L and Harouri, A and Lemaitre, A and others},
  journal={Phys. Rev. Lett.},
  volume={126},
  number={23},
  pages={233601},
  year={2021},
  publisher={APS}
}

@article{erhard2020advances,
  title={Advances in high-dimensional quantum entanglement},
  author={Erhard, Manuel and Krenn, Mario and Zeilinger, Anton},
  journal={Nat. Rev. Phys.},
  volume={2},
  number={7},
  pages={365--381},
  year={2020},
  publisher={Nature Publishing Group UK London}
}

@article{lodahl2017chiral,
  title={Chiral quantum optics},
  author={Lodahl, Peter and Mahmoodian, Sahand and Stobbe, S{\o}ren and Rauschenbeutel, Arno and Schneeweiss, Philipp and Volz, J{\"u}rgen and Pichler, Hannes and Zoller, Peter},
  journal={Nature},
  volume={541},
  number={7638},
  pages={473--480},
  year={2017},
  publisher={Nature Publishing Group UK London}
}

@article{chen2021bright,
  title={Bright solid-state sources for single photons with orbital angular momentum},
  author={Chen, Bo and Wei, Yuming and Zhao, Tianming and Liu, Shunfa and Su, Rongbin and Yao, Beimeng and Yu, Ying and Liu, Jin and Wang, Xuehua},
  journal={Nat. Nanotechnol.},
  volume={16},
  number={3},
  pages={302--307},
  year={2021},
  publisher={Nature Publishing Group UK London}
}
		
	\clearpage
	\newpage
	\onecolumngrid \bigskip

\end{document}